\documentclass[12pt]{iopart}

\RequirePackage{lineno}
\usepackage{graphicx}
\usepackage{graphics}
\usepackage{dcolumn}
\usepackage{epstopdf}
\usepackage{color}
\usepackage{amssymb}
\usepackage{bm}

\def\bea {\begin{eqnarray}}
\def\eea {\end{eqnarray}}

\def\be {\begin{equation}}
\def\ee {\end{equation}}

\begin{document}

\title{Study of jet-medium interactions using jet shape observables in heavy ion collisions at LHC energies with \textsc{JEWEL}}

\author{Rathijit Biswas$^{1}$, Subikash Choudhury$^{2}$, Sidharth Kumar Prasad$^{1}$ and Supriya Das$^{1}$  }
\address{$^{1}$ Center for Astroparticle Physics and Space Science \& Department of Physics, Bose Institute, 
Unified Academic Campus, EN-80, Sector-V, Bidhannagar, Kolkata - 91, India.}
\address{$^{2}$ Key Laboratory of Nuclear Physics and Ion-beam Application (MOE), and Institute of Modern Physics, Fudan University,
Shanghai-200433, People's Republic of China}
\ead{$^{1}$rathijit.biswas@cern.ch}
\ead{$^{2}$subikash@fudan.edu.cn}
\author{}

\vspace{10pt}

\date{\today}
\begin{abstract}
Based on a perturbative Quantum Chromodynamics (pQCD) inspired dynamical model of jet-medium interactions, \textsc{Jewel}, 
we have studied possible modifications to inclusive jet yields and a set of jet shape observables, namely, 
the fragmentation functions and radial momentum distributions when jets propagate through a deconfined partonic medium created 
in collisions of heavy nuclei at Large Hadron Collider (LHC) energies. Jets are reconstructed with anti-k$_{T}$ algorithm
in the pseudorapidity range $|\eta_{jet}| <$ 2.1 for resolution parameter \textit{R} = 0.2, 0.3 and 0.4. 
For background subtraction, a \textsc{Jewel}-compatible 4-Momenta subtraction technique (\textbf{4MomSub}) has been used. 
The modification of inclusive jet-yields in Pb-Pb collisions
relative to proton-proton interactions, quantified by R$_{AA}^{jet}$, are seen to be in reasonable agreement with
ALICE, ATLAS and CMS data over a broad transverse momentum range. \textsc{Jewel} is able to capture the qualitative features of the modifications to the 
fragmentation functions and radial momentum distributions in data but not always quantitatively. This quantitative discrepancy
may be related to the simplified treatment of recoil partons in the background model and partly due to background
subtraction procedure itself. Nevertheless, observed modifications to in \textsc{Jewel} corroborates the fact
that in-medium fragmentation is harder and more collimated than the fragmentation in vacuum. We further observe that these
modifications depend on the transverse momentum of jets and it seems that medium resolves the  core structure of low momentum jets
below 100 GeV/c at LHC energies.

\end{abstract}
\vspace{2pc}
\noindent{\it Keywords}: Jet-energy loss, Jet shapes, \textsc{Jewel}
\maketitle

\section{Introduction}
In relativistic collisions heavy nuclei (A-A), large suppression of single inclusive yields of high (transverse) momentum (p$_{T}$)
particles formed the basis for the landmark discovery of a new phase of QCD matter, characterized by a strongly interacting
but deconfined state of quarks and gluons (partons), namely the Quark Gluon Plasma (QGP) \cite{Intro_1, Intro_2}. It is known 
since long that the production of high-$p_{T}$ particles involve parton scatterings of large momentum transfer (hard scatterings), 
occurring at a time scale $<$ 1 fm/c \cite{Intro_3}. Each of these hard scattered partons then develop a shower of secondary partons through repeated gluon emission
and/or quark-antiquark splitting that eventually fragment into collimated spray of hadronic final states known as the Jet.
In a marked contrast to proton-proton (pp) interactions, jet production in heavy ion collisions is strongly suppressed, 
compatible with Bjorken-conjectured signature for the QGP formation, referred to as the “Jet quenching” \cite{Intro_4,
Intro_5, Intro_6, Intro_7, Intro_8, Intro_9}.

The jet-quenching  phenomena in its simplest form is mainfested as the suppression of the single inclusive cross-sections of
high-$p_{T}$ particles against a scaled pp reference at the same collision energy \cite{Intro_10, Intro_11, Intro_12, Intro_13, Intro_14}. 
This suppression, in general, encodes the
characteristic features of parton energy loss in a dense QGP, quantified by the nuclear modification factor or R$_{AA}$
\cite{Intro_15, Intro_16, Intro_16a, Intro_16b}.
But the measurements of R$_{AA}$ alone is not sufficient to decipher the complete picture of the parton energy loss 
mechanisms because of its insensitivity to the complex nature of parton shower evolution and the interaction between the
shower and medium constituents in QGP \cite{Intro_17}. 

This incompleteness can however, be alleviated from the measurements of fully reconstructed jets that are known to be a 
reasonable proxy to the shower initiating hard scattered partons. Jet being a composite object of hadronic final states 
(clustered together in a cone of radius, \textit{R}), it's energy loss has a subtle difference to the energy loss of an individual
parton. It is perceived that the modification of jet-yields in a medium not only depends on the energy loss of it's 
individual constituents, but also on how this lost energy is redistributed \cite{Intro_18}. 
The effective energy loss of jets in a medium
is therefore, consequences of the competing contributions of the jet-medium interactions causing the jet constituents to 
diffuse out-of the jet-cone leading to the degradation of its initial energy, while being compensated to a certain degree
by the in-cone emission of medium-induced gluons \cite{Intro_19, Intro_20, Intro_21, Intro_22}. 
Thus, the medium modifications of jets reflect the multifacet aspect of 
jet-medium interactions to which inclusive particle spectra are not at all sensitive. This substantiates on the necessity for 
studies with fully reconstructed jets in-order to examine the criticality of jet-medium interactions leading to the 
manifested features of jet-quenching phenomena: the suppression of jet-yields and the modification of intra-jet properties.

At Relativistic Heavy Ion Collider (RHIC) in BNL because of limited kinematic reach, measurements related to jet quenching were 
restricted to inclusive
hadron R$_{AA}$ or high-p$_{T}$ two-particle correlations \cite{Intro_23} (studies with fully reconstructed jets
are currently persued at RHIC as well \cite{RHIC_Jet_1, RHIC_Jet_2}).
At LHC, almost 10-fold increase in the collision energy resulted
in a significant rise in hard scattering processes, pushing the kinematic reach to several orders of magnitude. This has
largely facillitated studies with reconstructed jets. By now, we also have very rich theoretical insight on jet energy loss
mechanisms based on plethora of dynamical modelling of jet-medium interactions in an evolving QGP \cite{Jet_model_1,
Jet_model_2, Jet_model_3, Jet_model_4, Jet_model_5, Jet_model_6}. However, the multi-scale
nature of the problem makes the theoretical modelling of the jet-medium interactions highly non-trivial. To be a complete
theoretical framework, models need to consistently address different scales, ranging from the weakly interacting 
perturbative scale associated with initial hard scatterings and subsequent showering, to a non-perturbative scale of the
order-of $\Lambda_{QCD}$ when jets interact with the medium strongly. In addition, models need to incorporate jet-induced
medium response (recoil partons) that contribute to the so-called soft correlated background that can not be uniquely 
distinguished from the original jet-fragments.

To this end, Jet Evolution With Energy Loss model or \textsc{Jewel} \cite{Jet_model_3, JEWEL_1} 
has the necessary ingradients to demonstrate its effectiveness in describing variety of inclusive 
and differential jet observables. 
In this work we make an attempt to study different
aspects of the jet-quenching phenomena within \textsc{Jewel} and try to achieve further insight on the medium modification of fully
reconstructed jets via a systematic data-to-model comparison at the LHC energies. The organization of this paper is as 
follows. In section II an overview of \textsc{Jewel} and the parameter settings used for this work will be presented. Followed by 
results and discussions, and summary in section III and IV, respectively. 

\section{JEWEL}
\textsc{Jewel} offers a dynamical framework to simulate jet quenching phenomena in A-A collisions on the basis of QCD evolution of 
parton showers in presence of a thermalized partonic background. The initial state parton showers in \textsc{Jewel} are generated
with PYTHIA6.4 \cite{Pythia6} event generator on top-of \textsc{Cteq6LL} \cite{CTEQ6LL} parton distribution function (PDF) for pp events and EPS09
\cite{EPS09} nuclear PDF for heavy-ion collisions. A pp interaction in \textsc{Jewel} is identical with PYTHIA6, 
where as, for nuclear collisions PYTHIA generated parton
showers are coupled with Bjorken-type hydrodynamic background. The hydrodynamic expansion in \textsc{Jewel} 
follows an ideal-gas equation of state (EOS) with number and energy densities of medium partons being calculated taking local temperature in the medium as an input 
\cite{JEWEL_hydro_3}.
The in-medium scatterings of hard partons, resulting in elastic and in-elastic energy loss (QCD bremsstrahlung), are
described by 2$\rightarrow$n (n $\geq$ 2) leading order (LO) pQCD matrix elements protected against infra-red divergence. Radiations induced by 
re-scatterings are however, ordered according to its hardness, i.e, the parton shower will be perturbed by the hardest 
re-scattering \cite{Jet_model_3, JEWEL_scattering_1}. In addition, each re-scattering in \textsc{Jewel} is associated with a production of a recoil parton that propagates
through the medium without further interactions. Re-scatterings are terminated once the local temperature drops down to a
critical temperature T$_{c}$ (170 MeV), order of QCD phase transition temperature at baryon chemical potential $\mu_{B} \sim 0$.
For hadronization, recoil partons are first converted to gluons and then attached to strings connected with partons 
producing jet. This endows correlation between jets and recoil partons, likely to be manifested as enhancement in soft 
hadronic yields at large angles to jet-axis. Finally, PYTHIA takes over and hadronization occurs by Lund string 
fragmentation mechanism.

Our simulation results are based on standard \textsc{Jewel} setup in a ``recoil-on'' mode, i.e, keeping record of 
recoil partons in the event information. As recoil partons in \textsc{Jewel} retain their thermal component, i.e, momentum before
scattering, this essentially contributes to soft uncorrelated background specifically in the measurements of intra-jet 
observable and needed to be subtracted. Since \textsc{Jewel} does not simulate full event in A-A collisions, conventional background
subtraction procedures are rendered not suitable. Instead, \textsc{Jewel}-compatible background subtraction techniques were proposed
and tested over a variety jet-observables. We will use one of them, namely, the four momenta subtraction or 4MomSub
\cite{JEWEL_bkg_sub}.

For background subtraction in \textsc{Jewel} with 4MomSub, a set of neutral particles with very small momenta and positions 
corressponding to scattering centers, prior to interaction, are included in the final event information. These dummy 
particles together with others get clustured to a Jet. Then all the jet constituents pointing back to the scattering centers
(background) are identified and their four momenta are added. Finally, background four momenta is subtracted vectorially
from the four momenta of the reconstructed jets. This subtraction technique however, has certain limitations, for example,
it is only applicable to full jets and can not be used for charged jets directly \cite{JEWEL_bkg_sub} .

About one million 0-10\% central Pb-Pb and pp event sample were generated at $\sqrt{s_{NN}} =$ 2.76 TeV and 5.02 TeV.
The initial conditions required to simulate the hydrodynamic background, 
such as thermalization time $\tau_{i}$ and initial 
temperature T$_{i}$  are tabulated in TABLE~\ref{table:JEWELParam}. Jets are reconstructed with anti-k$_{T}$ \cite{antikt} jet-
reconstruction algorthim available in the \textsc{FastJet} package\cite{FastJet}.
\begin{table}[ht]
\caption{Details of the parameters used for JEWEL}
\centering
\begin{tabular}{|c|c|c|c|}
\hline 
$\surd s_{NN}$ & $\sigma_{NN}$ & $\tau_{i}$ & $T_{i}$ \\[0.5ex]
\hline
5.02 TeV & 72 mb & 0.4& 590 MeV\cite{JEWEL_hydro_1} \\
\hline
2.76 TeV & 64 mb & 0.6& 485 MeV\cite{JEWEL_hydro_2} \\
\hline
\end{tabular}
\label{table:JEWELParam}
\end{table}

\section{Results and Discussion}

\subsection{Jet nuclear modification factor (R$_{AA}^{jet}$)}
As a first step to characterize the effect of jet-medium interaction, we measure the nuclear modification factor for
single inclusive jet spectrum, defined as
\begin{equation}
R_{AA}^{jet} = \frac{d^{2}N^{jet}_{AA}/dp_{T,jet}d\eta}{<N_{coll}>d^{2}N_{pp}^{jet}/dp_{T,jet}d\eta},
\end{equation}
where N$^{jet}_{AA}$ and N$_{pp}^{jet}$ are the number of jets in A-A and pp collisions, respectively and 
$\langle N_{coll} \rangle$ is the number of binary nucleon-nucleon collision averaged over a given collison centrality
class. Figure 1 shows a comparison of $p_{T}$ dependence of $R_{AA}^{jet}$ calculated from \textsc{Jewel} with the recent
measurements from ALICE\cite{ALICE_2760, ALICE_5020}, ATLAS\cite{ATLAS_2760, ATLAS_5020} and 
CMS\cite{CMS_2760} for jet resolution parameter \textit{R} = 0.4 at $\sqrt{s_{NN}} =$ 5.02 TeV [Fig.1(a)] and 
2.76 TeV [Fig.1(b)]. 
\begin{figure}[h]
\centering
\includegraphics[scale = 0.44]{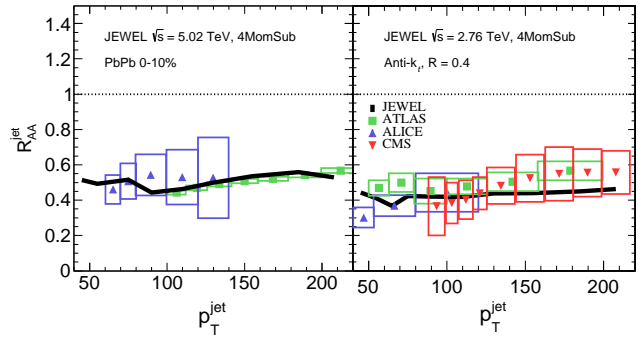}
\caption{[Color online] Transverse momentum dependence of R$_{AA}^{jet}$ from \textsc{Jewel} at \textit{R} = 0.4 in 0-10\% centrality, 
compared with ALICE \cite{ALICE_2760, ALICE_5020}, ATLAS \cite{ATLAS_2760, ATLAS_5020}
and CMS \cite{CMS_2760} measurements. \textbf{Left panel:} $\surd s$
= 5.02 TeV and \textbf{Right panel:} $\surd s$ = 2.76 TeV.}
\end{figure}
Data-to-model comparison shows a reasonable agreement, although, uncertainties in some p$_{T}$-bins in data are large. In A-A 
collisions, jets are usually reconstructed with small cone size (\textit{R}) to avoid large underlying background. This may lead to
additional suppression of jet cross sections because, partons scattered to large distances from jet-axis can not be 
recovered. This condition generally prevails when jets propagate through a medium. Energy and momentum transferred to medium
by jets are carried away to larger distances by medium expansion. This can influence the effective jet energy loss, leading
to a significant jet-cone size dependence. Also, the medium-stimulated radiations populating at large angles from the 
jet-axis can naturally account for additional reduction to jet energy. Expectedly, jets reconstructed in A-A collisions
with smaller cone size will capture only a subset of the full parton shower causing stronger attenuation of jet spectrum
relative to pp baseline. To illustrate the effect of \textit{R} dependence of jet cross sections in pp and A-A, the ratio of 
jet-p$_{T}$ spectra at \textit{R} = 0.2 with respect to different values of \textit{R} are shown in Fig.2.
\begin{figure}[h]
\centering
\includegraphics[width = 9 cm, height = 4.5 cm]{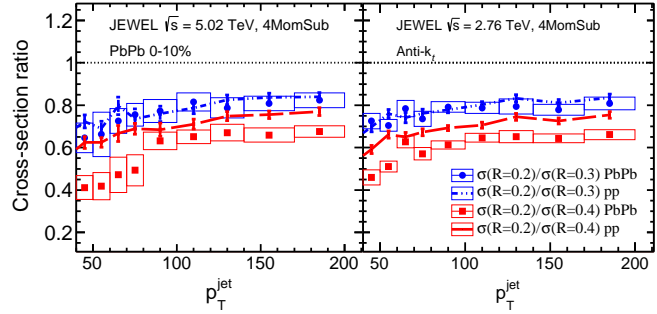}
\caption{[Color online] Ratios of jet p$_{T}$ spectra at \textit{R} = 0.3 and 0.4 with respect to \textit{R} = 0.2 in pp and 
Pb-Pb collisions. \textbf{Left panel:} $\surd s$ =
5.02 TeV and \textbf{Right panel:} $\surd s$ = 2.76 TeV}
\end{figure}
The ratio of cross sections exhibit a clear deviation from unity both in pp and Pb-Pb collisions, implying a large fraction
of initial partonic energy can be recovered by reconstructing jets with large \textit{R}. Interestingly, the ratio between \textit{R} = 0.3 to
\textit{R} = 0.2 in pp and Pb-Pb are consistent with each other, suggesting redistribution of energy by jet quenching does not
modify
the jet structure from \textit{R} = 0.2 to \textit{R} = 0.3. This could be because at these two radii we only observe the unresolved hard-core of
the entire jet. A significant difference between pp and Pb-Pb, beyond the statistical uncertainties, is however, observed
from \textit{R} = 0.2 to \textit{R} = 0.4. The R dependence of nuclear suppression factor has been measured by the CMS collaborations\cite{CMS_2760}
but given the current uncertainty in the measurements, suppression appears to be of \textit{R}-independent. Instead, some model calculations 
predict a clear hierarchy of R$_{AA}^{\mathrm{jet}}$ with \textit{R} \cite{LBT, model_PTSUM_1}. For completeness, we also calculate \textit{R} dependence of 
$R_{AA}^{jet}$ as function of p$_{T}$, shown
in Fig.3.
\begin{figure}[h]
\centering
\includegraphics[scale = 0.42]{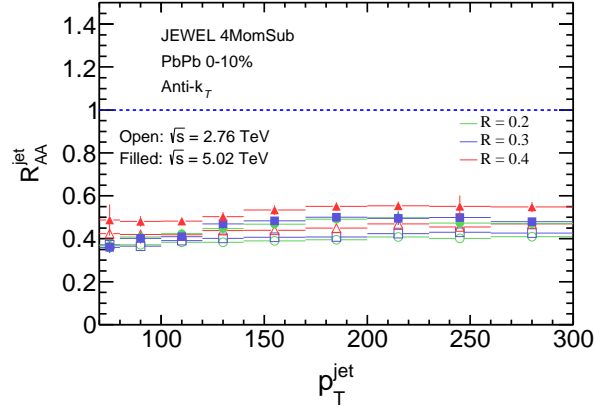}
\caption{[Color online] Transverse momentum dependence of R$^{jet}_{AA}$ at $\surd s$ = 5.02 and 2.76 TeV for resolution parameter \textit{R} = 0.2, 0.3 and 0.4 in
0-10\% central events in \textsc{Jewel}}
\end{figure}
We observe the level of suppression at \textit{R} = 0.2 and 0.3 is same, only difference is seen for \textit{R} = 0.4, consistent with our 
calculation of jet cross section ratios in Fig.2. To be mentioned, ratios of R$_{CP}$ ( central-to-peripheral jet cross
sections ratio), R$_{CP}^{0.3}$/R$_{CP}^{0.2}$, measured by the ATLAS collaboration \cite{ATLAS_RCP} also show similar 
trend for p$_{T} >$  100 GeV/c (ratio is seen to be consistent with unity). For direct comparison of R-dependence of jet suppression to data, 
$R_{AA}^{jet}$ for \textit{R} = 0.2 [Fig.4(a)] and \textit{R} = 0.4 [Fig.4(b)] from \textsc{Jewel} are overlaid on the CMS measurement \cite{CMS_2760}.
\begin{figure}[h]
\centering
\includegraphics[scale = 0.44]{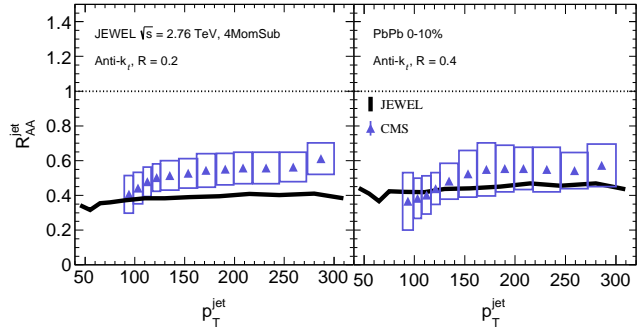}
\caption{[Color online] Transverse momentum dependence of R$_{AA}^{jet}$ from \textsc{Jewel} in 0-10\% centrality, compared with 
CMS \cite{CMS_2760} measurements. \textbf{Left panel:} \textit{R} = 0.2 and 
\textbf{Right panel:} \textit{R} = 0.4}
\end{figure}
\textsc{Jewel} underestimates data for \textit{R} = 0.2 but in case of \textit{R} = 0.4, agreement with data is better. A similar trend in data 
was also reported by the the ALICE Collaboration for charged-jets in \cite{ALICE_5020}. This may be because
the recoil partons in \textsc{Jewel} are scattered out to larger angles than the physical expectation. Hence, when \textit{R} is 
increased, recoil contributions are recovered leading to a better data-to-model agreement.

Besides the modification of inclusive jet yields from pp to A-A, jet-medium interactions also modify a number of intra-jet
properties, like, the angular and momentum distributions of jet-constituents within the cone. Modifications of these 
intra-jet properties can be examined from the measurements jet fragmentation functions and jet shapes. These are reported
in the sections to follow.

\subsection{Fragmentation function distributions(FF) }
As mentioned, the jet-quenching phenomena not only result in an overall jet energy loss but also modify the momentum 
distributions of particles within the jet. This modification can be quantified by measuring relative changes in the
fragmentation functions from pp to A-A collisions at same energy and a given interval of jet-p$_{T}$. The fragmentation 
function FF(z) \cite{ATLAS_FF_2760}, is generally defined as
\begin{equation}
FF(z) = \frac{1}{N_{jet}}\frac{dN}{dz},
\end{equation}
where \textit{z} = $\frac{ p_{T} \textrm{cos}\Delta R }{p_{T}^{jet}}$, refers to the longitudinal momentum fraction of 
particles relative to a jet of transverse momentum, p$_{T}^{jet}$. The quantity N$_{jet}$ denotes the number of jets in a
given momentum interval, and $\Delta R ( = \sqrt{ (\Delta\eta)^{2} + (\Delta\phi)^{2} } )$ corressponds to pseudorapidity
and azimuthal sepration of particles from the jet-axis.
\begin{figure}[h]
\centering
\includegraphics[scale = 0.44]{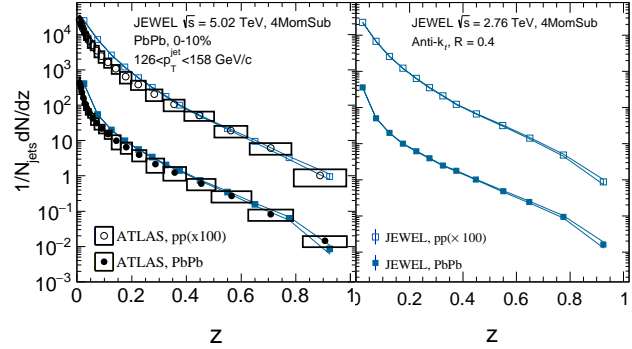}
\caption{[Color online] Fragmentation distributions from \textsc{Jewel} at \textit{R} = 0.4 in 0-10\% centrality compared with ATLAS \cite{ATLAS_FF_5020}
measurement of the same in pp and Pb-Pb 
collisions for 126$<p_{T}^{jet}<$158 GeV/c. \textbf{Left panel:} $\surd s$ = 5.02 TeV and \textbf{Right panel:} $\surd s$
= 2.76 TeV. Proton-proton data points are scaled by 100 for visibility. }
\end{figure}
Figure 5 shows FF(z) calculated within \textsc{Jewel} in pp and 0-10\% central Pb-Pb collisions at $\sqrt{s_{NN}}$ = 5.02
TeV in the p$_{T}^{jet}$ interval of 126 to 158 GeV/c. Also shown in the same, a quantitative comparison with the ATLAS
measurements from \cite{ATLAS_FF_2760}.
\begin{figure}[h]
\centering
\includegraphics[scale = 0.44]{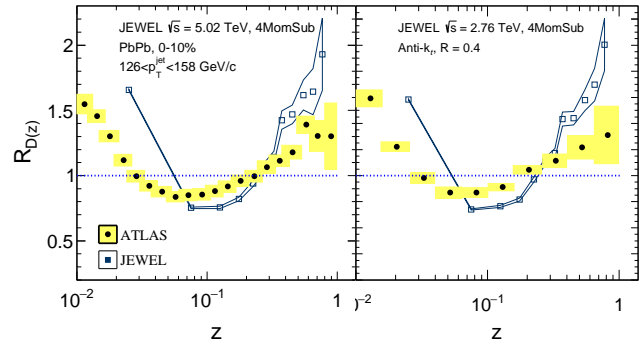}
\caption{[Color online] Ratio of fragmentation distributions from Pb-Pb to pp collisions from \textsc{Jewel} at \textit{R} = 0.4 in 0-10\% centrality compared with ATLAS \cite{ATLAS_FF_2760,ATLAS_FF_5020}
measurements in 126$<p_{T}^{jet}<$158 GeV/c. 
\textbf{Left panel:} $\surd s$ = 5.02 TeV and \textbf{Right panel:} $\surd s$ = 2.76 TeV. }
\end{figure}
Furthermore, to quantify the modification of fragmentation functions in heavy ion collisons, ratios of FF(z) in Pb-Pb 
to that in pp
collisions are shown in Fig.6 and compared with the ATLAS data for collision energies 5.02 TeV and 2.76 TeV 
\cite{ATLAS_FF_2760, ATLAS_FF_5020}. The shape of
the distributions from the model calculation are similar to data, with following generic features: an enhancement at low
and high z and a depletion at intermediate z. The enhancement at low z is interpreted as the softening of the 
fragmentation functions, i.e, an excess yield of low-p$_{T}$ particles within the jet-cone. 

While some model calculations
attribute this excess to multiple soft gluon emissions, in \textsc{Jewel}, enhancement is mainly from the jet-induced 
medium response. It is realized, because of the momentum conservation, jet-energy deposited into the medium will have a 
component along the jet direction ( unless completely thermalized ). As a result, particles produced from the medium in the
vicinity of jets will be endowed with kinematic focussing along the jet-axis. These particles being reconstructed as a part 
of a jet will soften the momentum distributions of hadronic fragments of jets in a medium than in vacuum. Quantitatively,
\textsc{Jewel} calculation overshoots the data at low-z. This discrepency may be related to the fact that recoil partons in
\textsc{Jewel} suffer only single interaction. Therefore, the degree of correlation between jet and soft recoils could be 
larger than expected in a physical scenario where recoils suffer multiple interactions and tend to thermalize. Here we would like
to mention that the background subtraction in \textsc{Jewel} is done at the inclusive level and not at the track level this might
contribute to the discrepency between data and \textsc{Jewel} at the level of a few percent \cite{JEWEL_bkg_sub}.

On the other hand, observed enhancement at high-z is attributed to hardening and narrowing of jets because of the energy
loss. A possible explanation to this was offered by the apparent similarity in the fragmentation functions of quenched jets
with the jets initiated by quarks in vacuum. A quark-initiated jet is in general, harder and more collimated \cite{ALICE_JetShape} . 

Similar enhancement at high-z is also expected from the color coherence effect\cite{CC_1, CC_2}.
Partonic fragments with transverse spread less than the
coherent length scale of the medium, remain unresolved and shielded against the medium modification. As a consequence, 
although a quenched jet has a overall energy less than the jet in vacuum at same p$_{T}$ but it may have originated from a
more energetic parton. Since a sizeable fraction of the total jet energy is concentrated within the unresolved hard core, 
despite the energy loss, modification to intra-jet structures are negligible. This may actually bias an ensemble of 
quenched jets in A-A collisions to higher virtuality relative to pp when compared at same jet-p$_{T}$. In \textsc{Jewel} a
similar situation may have been mimicked because the re-scatterings of shower partons with medium are ordered according to
virtuality. Thus, a scattering harder than the virtuality of the shower parton would result in a medium-induced radiation.
Since the re-scatterings are softer than the orginal scale of the parton shower, hard inner core remain unmodified, leading
to the observed enhancement in fragmentation function ratios at high-z.
\begin{figure}[h]
\centering
\includegraphics[scale = 0.44]{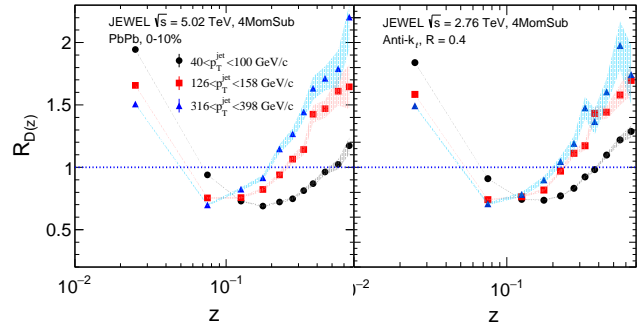}
\caption{[Color online] Ratios of fragmentation distributions from Pb-Pb to pp collisions from \textsc{Jewel} at \textit{R} = 0.4 in 0-10\% centrality for
three p$_{T}^{jet}$ ranges: 40$<p_{T}^{jet}<$100, 126$<p_{T}^{jet}<$158 
and 316$<p_{T}^{jet}<$398 GeV/c. \textbf{Left panel:} $\surd s$ = 5.02 TeV and \textbf{Right panel:} $\surd s$ = 2.76
TeV.}
\end{figure}

To further determine how the modification of fragmentation functions depends on p$_{T}^{jet}$, ratios of fragmentation functions in
three jet-p$_{T}$ intervals: 40 \textless p$_{T}^{jet}$ \textless 100 GeV/c, 126 \textless p$_{T}^{jet}$ \textless 158
GeV/c and 316 \textless p$_{T}^{jet}$ \textless 398 GeV/c are compared in Fig.7. A clear excess of soft particles is 
observed in all jet-p$_{T}$ intervals together with an ordering. While low-z excess is maximum in the lowest  p$_{T}^{jet}$
interval, the high-z enhancement in the same p$_{T}^{jet}$ range is seen to be least. This might indicate a significant 
modification to jet substructures at this low p$_{T}^{jet}$ interval. With high statistics data being available
from the LHC Run II, such predictions can be tested to achieve insight on the transverse resolution scale of 
the medium at the LHC kinematics.

Next we move to study a classic jet shape observable that has been designed to analyse the radial distribution
of the jet transverse momentum from the jet axis.

\subsection{Radial distribution of transverse momentum density}
The radial distribution of transverse momentum density, dp$_{T}$$^{sum}$/dr, within a jet is measured as a function of the
distance \textit{r} = $\sqrt{(\Delta\eta)^{2} + (\Delta\phi)^{2}}$ radially outword from the jet axis. The momentum density 
\begin{equation}
  <\frac{dp_T^{sum}}{dr}>(r) = \frac{1}{\Delta r} \frac{1}{N_{jets}}\sum^{N_{tracks}}_{i = 1}p_{T}^{i}(r - \Delta r/2,r + \Delta r/2)
\end{equation}
is calculated by taking scalar sum of the transverse momentum, p$_{T}^{sum}$, of all particles that falls within the 
annular regions of width $\Delta$r at radius \textit{r} \cite{CMS_PTSUM}.

\begin{figure}[h]
\centering
\includegraphics[scale = 0.44]{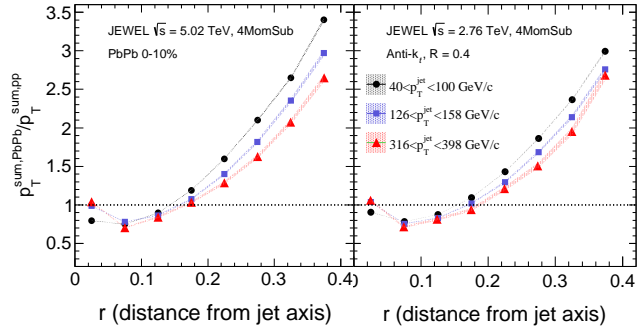}
\caption{[Color online] Ratios of radial momentum density distributions from PbPb to pp collisions from \textsc{Jewel} at \textit{R} = 0.4 in 0-10\% centrality
 are presented for 40$<p_{T}^{jet}<$100, 126$<p_{T}^{jet}<$158 GeV/c and 316$<p_{T}^{jet}<$398 p$_{T}^{jet}$ ranges. \textbf{Left panel:} $\surd s$ = 5.02 TeV
and \textbf{Right panel:} $\surd s$ = 2.76 TeV}
\end{figure}
Figure 8 shows the modification in jet shape distribution in 0-10\% central Pb-Pb collisions relative to pp baseline.
The ratio of jet shape distribution functions in A-A to pp collision is seen to have depletion at intermediate \textit{r} and an
enhancement at large \textit{r}. However, at small \textit{r}, jets with p$_{T}^{jet}$ \textgreater 100 GeV/c modification factor is 
consistent with unity, implying that jet core remains unmodified or negligibly modified.

While some model calculations attribute this enhancement at large-\textit{r} to wide-angle medium induced gluon radiation
and/or momentum broadening \cite{Intro_21}, but, in \textsc{Jewel}, 
enhancement at large-\textit{r} can only be accounted on inclusion of the recoil partons. 
Interestingly, in the jet-p$_{T}$ range 40 to 100 GeV/c, depletion persists down to very small \textit{r}.
This is consistent with what we inferred from the modification of fragmentation functions of jets \textless 100 GeV/c. That
is, low momentum jets, preferably \textless 100 GeV/c at LHC energies might have their inner core modified by the 
jet-medium interaction.
\begin{figure}[h]
\centering
\includegraphics[scale = 0.44]{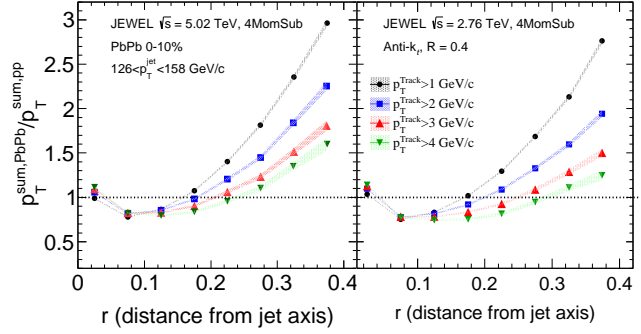}
\caption{[Color online] Ratios of radial momentum density distributions from PbPb to pp collisions from \textsc{Jewel}
at \textit{R} = 0.4 in 0-10\% centrality for 126$<p_{T}^{jet}<$158 GeV/c, in different ranges of constituent particle's  p$_{T}$. 
\textbf{Left panel:} $\surd s$ = 5.02 TeV and \textbf{Right panel:} $\surd s$ = 2.76 TeV}
\end{figure}

We further decompose the excess at large-\textit{r} into different ranges of constituent particle's p$_{T}$ inside the jet cone.
Here in Fig.9, we note that the excess is more pronounced in the lowest p$_{T}$ range, 1 \textless p$_{T}$ \textless 2 GeV/c,
which is consistent with the enhancement seen in the measurements of jet-track correlations in \cite{CMS_JetTrack_Corr}. 
This certainly suggests that a sizeable fraction of total jet-energy is carried out to large relative angles from the jet axes
and concentrated in the low p$_{T}$ ranges of the overall spectrum.

\section{Summary}
 Manifestations of jet-medium interactions are not only limited to the suppression of single inclusive jet spectrum but
also reflected in the modifications of it's transverse and longitudinal energy or momentum profile. 
In this work we have investigated the effect of jet energy loss on the single inclusive jet yields and 
intra-jet properties, namely, the jet fragmentation functions and jet shapes, within a dynamical framework of jet-medium interaction, 
\textsc{Jewel}, at LHC energies. \textsc{Jewel} calculations with appropriately chosen parameters (constrained to describe bulk obsevables
in high energy heavy ion collisions) can reasonably describe the p$_{T}$ dependence of jet yield modification in data when
evaluated at jet resolution parameter, \textit{R} = 0.4, as shown in Fig.1.
We have repeated the same exercise for smaller values of \textit{R}, as jet-medium interactions are likely to redistribute the jet energy over a 
wide angular spread. Therefore, one may expect stronger suppression of jet spectrum when calculated at smaller values of \textit{R}.
Indeed such an effect has been observed in Fig.2 where the jet cross-section ratios, $\sigma(\textit{R} = 0.2)/\sigma(\textit{R} = 0.3)$ 
in Pb-Pb and pp are seen to deviate from unity but their magnitudes are consistent. 
A significant difference between Pb-Pb and pp is however, noticed for $\sigma(\textit{R} = 0.2)/\sigma(\textit{R} = 0.4)$. This 
essentially suggests that dynamics of jet-medium interaction in 
\textsc{Jewel} is able 
to capture the large angle transport of jet energy in medium which is a key feature responsible for the jet energy loss.
Same conclusion can also be drawn from Fig.3 and Fig.4 where jet suppression factor is seen to exhibit observable 
difference when evaluated at small (0.2) and large (0.4) \textit{R}.

Further details on the jet-medium interactions can be extracted from measurements of jet fragmentation functions and jet shapes.
We observe quenched jets have a harder fragmentation pattern than jets in vacuum when compared in the same jet-p$_{T}$ interval.
In addition a p$_{T}$-dependent modification to these fragmentation functions are also observed. Modification quantified
by the ratios of the fragmentation functions in A-A collisions to those in pp exhibit large enhancement at low and high z 
($\sim$ p$_{T}$) as shown in Fig.7.
This excess yield of soft fragments is generally attributed to wide angle emission of medium-induced soft gluons but in 
\textsc{Jewel} soft excess could only be achieved on inclusion of recoil contributions.
The enhancement of hard fragments (jet-constituents with relatively large
transverse momentum and generally aligned to the jet axis), on the other hand, could be because of quark-like fragmentation pattern on top-of
the hardening of the initial energy spectrum of quenched jets when compared to vacuum-jets in the same jet-p$_{T}$ interval. 
It is worthy to note that in the jet-$p_{T}$ range 40 to 100 GeV/c,
enhancement at high-z drops significantly. This may be because at this jet-p$_{T}$ interval, jet-medium interactions are strong enough
to resolve the inner hard core of jets. Similar conclusions has been reached from the measurements complementary to
the fragmentation functions i.e jet shapes, presented in Fig.8.

\medskip

\noindent
{\bf Acknowledgement}

Significant part of computation for this work was carried out using the computing server facility at CAPSS, 
Bose Insititute, Kolkata. RB is supported by DST sponsored project SR/MF/PS-01/2014-BI. 
SC acknowledges financial support granted under SERB (DST) sponsored ``Ramanujan Fellowship``
of Dr. Saikat Biswas (D.O. no. SR/S2/RJN-02/2012). RB and SC would like to acknowledge fruitful discussions with 
Raghav Kunnawalkam Elayavalli on background subtraction in \textsc{Jewel}.

\end{document}